\documentclass[pre,twocolumn,superscriptaddress,amsmath,amssymb,nofootinbib]{revtex4-2}

\usepackage{graphicx}
\usepackage{bm}
\usepackage{xcolor}
\usepackage[colorlinks=true,linkcolor=blue,citecolor=blue,urlcolor=blue]{hyperref}

\newcommand{\dd}{\mathrm{d}}
\newcommand{\pa}{\partial}
\newcommand{\La}{\mathcal{L}_a}
\newcommand{\phat}{\hat{p}}

\newcommand{\pitfall}[1]{\par #1\par}

\begin{document}

\raggedbottom  

\title{Numerical integration of non-local master equations for renewal shot noise
with drift and state-dependent amplitude}

\author{M.~Bianucci}
\affiliation{Istituto di Scienze Marine, Consiglio Nazionale delle Ricerche (ISMAR--CNR), 19032 Lerici (SP), Italy}
\author{R.~Mannella}
\affiliation{Dipartimento di Fisica, Universit\`a di Pisa, 56100 Pisa, Italy}

\date{\today}

\begin{abstract}
The probability density of a system driven by a drift and by a renewal train of kicks with
state-dependent amplitude obeys a master equation that is non-local in space, because the kick
operator $\phat(i\pa_xI(x))$ is of infinite order, and, for non-exponential waiting times,
non-local in time, with a memory kernel that contains the propagator of the drift. Its
solutions contain quasi-singular components, the mass that has not yet been kicked, which
collapse onto the attractors of the drift. We present a numerical method that integrates such
equations accurately. The kick operator is a push-forward along the Marcus flow;
in the Lamperti variable it becomes a translation, and its exact cell-to-cell weights follow
from the cumulative distribution of the kicks, without quadrature. The drift is integrated by
a conservative semi-Lagrangian step that is exact in time. The memory is represented by a sum
of exponentials whose auxiliary fields are carried by the drift, at a cost linear in the number
of time steps. The never-kicked mass is transported exactly. For each ingredient we quantify
the errors introduced by simpler alternatives. The method is verified against exact solutions
and against event-driven Monte Carlo simulations with exact flow maps: with the never-kicked
mass treated exactly, it is statistically indistinguishable from $10^8$ trajectories. The
numerical parameters determined for Gaussian kicks carry over unchanged to other kick
distributions, and near a zero of the coupling, where the density vanishes faster than any power,
the solution follows the known asymptotic law down to probability densities of order $10^{-37}$,
far beyond the reach of Monte Carlo sampling. As an application we compare the exact equation
with its local and Poissonian approximations:  in the range explored, the local equation
deviates by up to 8\% of the width of the density, also for waiting-time distributions with
finite mean, consistently with an estimate of its error that predicts larger deviations for
slow drifts at long times. A Fortran
implementation is available.
\end{abstract}

\maketitle

\section{Introduction}
\label{sec:intro}

Impulsive forcing with state-dependent gain is common in physics and beyond: synaptic
conductance input, intermittent forcing of slow climate modes, crack growth, transport by
discrete collisions. Its simplest description is
\begin{equation}
\dot x=-C(x)+I(x)\,\xi(t),\qquad \xi(t)=\sum_k\xi_k\,\delta(t-t_k),
\label{eq:sde}
\end{equation}
where the amplitudes $\xi_k$ are i.i.d.\ with density $p(\xi)$ and the event times $t_k$ form a
renewal process with waiting-time density (WTD) $\psi(t)$. For impulses of short but finite
duration the kick is naturally read in the Marcus sense
\cite{Marcus1978,Marcus1981,ChechkinPavlyukevich2014,BianucciMannellaFNL}: a kick of amplitude
$\xi$ moves the state along the flow of $\dd x/\dd s=I(x)$ for a virtual time $\xi$.
(Reference~\cite{BianucciMannellaFNL} writes the coupling with the opposite sign , and so does
Ref.~\cite{Bianucci2026}, $\dot x=-C-I\xi$, whose kick operator, as corrected in
Ref.~\cite{Bianucci2026corr}, is $\phat(-i\pa_xI)$; the forms differ by $\xi\to-\xi$.)

The evolution of the probability density $P(x,t)$ of Eq.~\eqref{eq:sde} was derived exactly in
Ref.~\cite{Bianucci2026}, together with a local approximation, the universal local master
equation (ULME) . The ULME is exact for Poissonian events and, in general, to first order in the
kick operator, but it is not a long-time limit  of the exact equation \cite{Bianucci2026corr}.
The numerical integration of these equations raises three difficulties.
(i) The kick operator $\phat(i\pa_xI)$, with $\phat$ the characteristic function of $p(\xi)$,
is of infinite order. Truncating its Kramers--Moyal expansion is not a controlled approximation
when the kicks are finite: the series in the kick amplitude need not converge usefully
\cite{DiPaolaFalsone1993,BianucciMannellaFNL}, and it fails altogether for heavy-tailed kicks.
(ii) For non-exponential waiting times the equation is non-local in time, and its memory
kernel contains the free propagator of the drift: the system is transported between kicks.
Neglecting this propagator changes the equation, while storing the full history makes the cost
grow quadratically with the number of time steps. An exact Markovian alternative adds the age of
the renewal process as a further variable \cite{Cox1962}, at the price of one more dimension; the
auxiliary fields of Sec.~\ref{sec:soe} can be read as a compressed representation of that
dimension.
(iii) The mass that has not yet been kicked is transported by the drift alone and collapses onto
its attractors. For power-law WTDs this mass decays slowly, so that the density carries, for
long times, components narrower than any fixed grid.

Equations of this type have been treated in several communities: Fokker--Planck equations for
Marcus SDEs \cite{DenisovHorsthemkeHanggi2009,SunLiZheng2014}, the Kolmogorov--Feller equation
with parametric Poissonian impulses \cite{DiMatteoDiPaolaPirrotta}, population-density
methods for neurons with non-Poissonian input \cite{LaiDeKamps2016}, jump--diffusion
equations in finance \cite{dHalluin2005}. These approaches address subsets of the difficulties
above: Poissonian events, additive jumps, or memory kernels evaluated over the stored history.
Here we present a method for the general case of non-Poissonian renewal events, nonlinear drift,
state-dependent kicks and arbitrary kick distributions, and we verify it against exact solutions
and Monte Carlo simulations. It extends and supersedes the numerical section of
Ref.~\cite{Bianucci2026}.

Section~\ref{sec:equations} recalls the equations. Section~\ref{sec:blocks} describes the
treatment of the kick operator, the drift, the memory and the never-kicked mass, comparing each
choice with simpler alternatives, and the resulting algorithm. Section~\ref{sec:verification}
presents the verification: exact solutions, Monte Carlo simulations up to $10^8$ trajectories,
other kick distributions, and the approach to a zero of $I$, where the density is compared with
an asymptotic law over more than thirty decades. Section~\ref{sec:examples} compares the exact
equation with its local and Poissonian approximations, and Sec.~\ref{sec:recommendations}
summarises the numerical parameters required in the cases studied.

\section{The equations}
\label{sec:equations}

\emph{Kicks.} Let $\phi_s$ be the flow of $\dd x/\dd s=I(x)$ and $G(x)=\int^x\dd z/I(z)$ the
Lamperti variable, so that $\phi_s(x)=G^{-1}(G(x)+s)$: in $G$ the flow is a translation. A
Marcus kick maps $x\mapsto\phi_{\xi}(x)$. On densities it acts as
\begin{equation}
\left(e^{-\xi\pa_xI}P\right)(x)=P\big(\phi_{-\xi}(x)\big)\,\frac{I(\phi_{-\xi}(x))}{I(x)},
\label{eq:pushforward}
\end{equation}
the push-forward of $P$ through $\phi_{\xi}$, and the averaged kick operator is
$\phat(i\pa_xI)=\int p(\xi)\,e^{-\xi\pa_xI}\,\dd\xi$. Zeros of $I$ are fixed points of every
kick and cannot be crossed by any of them \cite{BianucciMannellaFNL}: if $I(x^*)=0$ and $I$ is
Lipschitz, $|G(x)|\to\infty$ as $x\to x^*$, so that $G(x)+\xi$ stays finite for any finite $\xi$
and $\phi_\xi$ maps each interval on which $I$ has constant sign into itself.

\emph{Renewal process.} With survival function $\Psi(t)=\int_t^\infty\psi(s)\,\dd s$, the renewal density
$R(t)$ (rate of events at time $t$ given an event at $t=0$) solves the renewal
equation~\cite{Cox1962}
\begin{equation}
R(t)=\psi(t)+\int_0^t\psi(t-u)\,R(u)\,\dd u .
\label{eq:volterra}
\end{equation}
In the examples we use the Pareto WTD $\psi(t)=\frac{\mu-1}{T}(1+t/T)^{-\mu}$, $\mu>1$.

\emph{Master equations.} With the drift operator $\La P=\pa_x[C(x)P]$, all equations considered
here have the form
\begin{equation}
\pa_tP=\La P+\left[\phat(i\pa_xI)-1\right]\Phi[P](x,t),
\label{eq:me}
\end{equation}
and differ in the ``kick flux'' $\Phi$:
\begin{align}
\text{exact:}\quad &\Phi=R(0)P(x,t)\nonumber\\
&\quad+\int_0^tR'(t-u)\,e^{(t-u)\La}P(x,u)\,\dd u, \label{eq:exact}\\
\text{ULME:}\quad &\Phi=R(t)\,P(x,t), \label{eq:ulme}\\
\text{Poisson:}\quad &\Phi=P(x,t)/\langle\tau\rangle . \label{eq:poisson}
\end{align}
Equation~\eqref{eq:exact} is exact for any renewal process started with an event at $t=0$
\cite{Bianucci2026}; Eq.~\eqref{eq:ulme} is its local approximation, exact to first order in
$\phat-1$ \cite{Bianucci2026corr}; Eq.~\eqref{eq:poisson}
is exact for exponential waiting times, for which $R'\equiv0$ and the three coincide.

\emph{Renewal representation.} The kick flux $\Phi$ has a direct meaning: $\Phi(x,t)\,\dd x\,\dd t$
is the probability that a kick occurs in $[t,t+\dd t]$ with the system in $[x,x+\dd x]$ just before
it. Let $U(t,u)$ be the propagator between kicks, here $e^{(t-u)\La}$, and $K=\phat(i\pa_xI)$ the
kick operator. For a renewal process started with an event at $t=0$,
\begin{align}
P(t)&=\Psi(t)\,U(t,0)P_0+\int_0^t\Psi(t-u)\,U(t,u)K\Phi(u)\,\dd u,\nonumber\\
\Phi(t)&=\psi(t)\,U(t,0)P_0+\int_0^t\psi(t-u)\,U(t,u)K\Phi(u)\,\dd u :
\label{eq:flux}
\end{align}
a state at time $t$ has either never been kicked, or was last kicked at time $u$ and has evolved
freely since then. The renewal identity $R(0)\Psi(t)+\int_0^tR'(t-u)\Psi(u)\,\dd u=\psi(t)$, i.e.\
$s\hat R\hat\Psi=\hat\psi$ in Laplace space, and the semigroup property $U(t,u)U(u,v)=U(t,v)$
give $\Phi=R(0)P(t)+\int_0^tR'(t-u)\,U(t,u)P(u)\,\dd u$, which is Eq.~\eqref{eq:exact}, and
differentiating the first of Eqs.~\eqref{eq:flux} gives Eq.~\eqref{eq:me}. The derivation uses
only the semigroup property of $U$ and the time independence of $K$; the scheme of
Sec.~\ref{sec:blocks} rests on the same two properties.

\emph{The never-kicked mass.} A fraction $\Psi(t)$ of the probability has not been kicked by
time $t$; it is the initial density transported by the drift. For the Pareto WTD it decays as
$(1+t/T)^{1-\mu}$, i.e., slowly. Near a stable fixed point $x^*$ of $-C$ its width shrinks as
$e^{-|C'(x^*)|t}$. This component is the main source of numerical difficulty
(Sec.~\ref{sec:split}).

\emph{Test cases.} Two systems are used throughout, both with Pareto waiting times of scale $T=1$
and, unless stated otherwise, Gaussian kicks of unit variance:
\begin{itemize}
\item[(A)] linear drift and coupling, $C=\gamma x$, $I=1+0.5x$, with a single attractor $x^*=0$ and
a zero of $I$ at $x=-2$;
\item[(B)] bistable drift, $C=\gamma(-1.2x+x^3)$, $I=1+0.3x$, with attractors $x^*=\pm\sqrt{1.2}$.
\end{itemize}
The values of $\gamma$, $\mu$ and of the final time are given where each case is used.

\section{Building blocks}
\label{sec:blocks}

We use a uniform finite-volume grid of $N_x$ cells of width $\Delta x$ on $[x_{\min},x_{\max}]$,
cell centres $x_j$, faces $x_{j\pm1/2}$, and Strang splitting of Eq.~\eqref{eq:me}: half a step
of drift, one step of kicks, half a step of drift.

\subsection{The kick operator as a push-forward}
\label{sec:kernel}

By Eq.~\eqref{eq:pushforward}, the mass located at $x_j$ is moved by a kick $\xi$ to
$\phi_{\xi}(x_j)$, which lies in cell $i$ if and only if $G(x_j)+\xi\in[G_{i-1/2},G_{i+1/2}]$,
with $G_{i\pm1/2}=G(x_{i\pm1/2})$ (for $I>0$). The exact transition weights are therefore
\begin{equation}
K_{ij}=F_\xi\!\big(G_{i+\frac{1}{2}}-G_j\big)-F_\xi\!\big(G_{i-\frac{1}{2}}-G_j\big),
\label{eq:kernel}
\end{equation}
with $F_\xi$ the cumulative distribution of the kicks and $G_j=G(x_j)$, and
$[\phat(i\pa_xI)\Phi]_i\simeq\sum_jK_{ij}\Phi_j$. The matrix is assembled once and stored in
sparse form. It is non-negative; its column sums equal one minus the probability of leaving the
domain, which is recorded. $G$ is analytic for linear $I$ and is computed
by Gauss--Legendre quadrature otherwise. At a zero of $I$, $G\to\mp\infty$, and the barrier of
Ref.~\cite{BianucciMannellaFNL} is respected exactly. For discrete kick distributions
(dichotomous kicks) the images $\phi_{\xi}(x_j)$ are deposited on the two nearest cell centres
with linear weights.

\pitfall{A direct discretisation of $\int p(\xi)e^{-\xi\pa_xI}\dd\xi$ by a Gauss rule in $\xi$
(Hermite for Gaussian kicks, Laguerre for exponential ones) converges slowly when
$\Phi$ contains structures narrower than the spacing of the nodes, which is precisely the case
near attractors. On an exact benchmark, Gauss rules reach the accuracy of the cell-integrated
kernel~\eqref{eq:kernel} only with 60--120 nodes for Gaussian kicks and 120--240 for Laplace and
one-sided exponential kicks, whereas Eq.~\eqref{eq:kernel} requires no such choice
(Appendix~\ref{app:kernel}).
A truncated Kramers--Moyal expansion is not a controlled approximation for finite kicks.}

\subsection{The drift: conservative semi-Lagrangian step}
\label{sec:csl}

For the drift substep we transport the cell faces with the exact flow of $\dot x=-C(x)$
(analytic for linear $C$, adaptive Runge--Kutta otherwise) and set
\begin{equation}
P_i^{\rm new}=\frac{M\big(X_{i+1/2}\big)-M\big(X_{i-1/2}\big)}{\Delta x},
\label{eq:csl}
\end{equation}
where $X_{i\pm1/2}$ are the departure points of the faces and $M$ is the cumulative mass of a
piecewise-linear, slope-limited reconstruction \cite{Crouseilles2010}. The scheme conserves
mass and positivity, is not subject to the Courant--Friedrichs--Lewy (CFL) limit
$|C|\Delta t\lesssim\Delta x$ of explicit schemes and, for autonomous $C$, has no
time-discretisation error; its spatial error is second order.

\pitfall{Flux-limited Lax--Wendroff schemes are frequently used for the advection term. If the
second-order correction is built on $P$ rather than on the flux $vP$, with $v=-C$, the term
$\tfrac12\Delta t\,v v'P$ is lost and the scheme is first order in time whenever $C$ is not
constant; with the correction built on the flux, the error of a benchmark drops by a factor of
20 at the same resolution.}

\subsection{The memory: sum of exponentials carried by the drift}
\label{sec:soe}

$R$ is obtained from Eq.~\eqref{eq:volterra} by implicit trapezoidal quadrature, and $R'$ is
approximated by a sum of exponentials, $R'(u)\simeq\sum_{m=1}^Mc_me^{-\lambda_mu}$, with
log-spaced rates and weighted least squares \cite{Jiang2017,Lubich1988} (Appendix~\ref{app:soe}). Then
$\Phi=R(0)P+\sum_mY_m$, with
\begin{equation}
\pa_tY_m=\La Y_m-\lambda_mY_m+c_mP,\qquad Y_m(0)=0,
\label{eq:soe}
\end{equation}
which is equivalent to Eq.~\eqref{eq:exact} for the fitted kernel. The drift part of
Eq.~\eqref{eq:soe} is applied to the $Y_m$ in the same substeps as to $P$, so that the
propagator $e^{(t-u)\La}$ is included without storing the history; the relaxation and source
terms are integrated exactly over each step (Appendix~\ref{app:soe}). The cost per step is
$O(MN_x)$ instead of $O(N_tN_x)$. With $M=40$ the relative fit error on $R'$ is
$\simeq3\times10^{-9}$, and increasing $M$ to 60 changes nothing visible even against $10^8$
trajectories (Sec.~\ref{sec:stress}).

 \pitfall{Replacing $e^{(t-u)\La}$ by the identity, i.e., neglecting the motion of the system
during the waiting times, gives the Montroll--Weiss memory term \cite{MontrollWeiss1965}
supplemented by the drift; this is exact only for $C\equiv0$. With $C=x$, $I=1+0.5x$ and
$\mu=1.5$ this model deviates from a Monte Carlo sample of $N=2\times10^6$ trajectories by
$5.2\,\sigma_{\rm MC}$ (rms over histogram bins, in units of the statistical error of each bin; see
Sec.~\ref{sec:mc}) already at $t=1.2$, twice the deviation of the local
approximation~\eqref{eq:ulme}. The analogous question, how
an external force enters the memory operator, arises for fractional Fokker--Planck
equations~\cite{MetzlerBarkaiKlafter1999,SokolovKlafter2006,Heinsalu2007,HenryLanglandsStraka2010}.}

\subsection{The never-kicked mass}
\label{sec:split}

The never-kicked component is known exactly: it is the initial density transported by the
drift, with weight $S(t)$ equal to $\Psi(t)$ for Eq.~\eqref{eq:exact},
$\exp[-\int_0^tR(s)\,\dd s]$ for Eq.~\eqref{eq:ulme} and $e^{-t/\langle\tau\rangle}$ for
Eq.~\eqref{eq:poisson}. We therefore write
\begin{equation}
P(x,t)=S(t)\,(\varphi_t)_\#P_0(x)+Q(x,t),
\label{eq:split}
\end{equation}
where $\varphi_t$ is the flow of the drift. The first term is evaluated exactly, by flowing the
faces backward and integrating $P_0$ between them. $Q$, with $Q(x,0)=0$, obeys the same equation
as $P$ with an additional source, the rate of first kicks $-\dot S$ times
$\phat(i\pa_xI)(\varphi_t)_\#P_0$. For Eq.~\eqref{eq:exact} this follows from the renewal identity of Sec.~\ref{sec:equations},
together with $e^{(t-u)\La}(\varphi_u)_\#P_0=(\varphi_t)_\#P_0$; accordingly, the
auxiliary fields of Eq.~\eqref{eq:soe} are driven by $Q$ rather than by $P$.
The quasi-singular component never lives on the grid. The same splitting treats exactly a
process observed at an age $t_a>0$: $\Psi$ is replaced by the survival function $\Psi_1$ of the
first event after the start of the observation, and the source by its density $\psi_1$, while
the memory acting on $Q$ remains the non-aged one \cite{Bianucci2026corr}.

\pitfall{Without Eq.~\eqref{eq:split}, the never-kicked mass collapses onto the attractors to
widths far below $\Delta x$ and dominates the error (Table~\ref{tab:stress}). A second effect
remains with it: if an attractor lies exactly on a cell face, the round-off of the face
position is amplified as $e^{|C'(x^*)|t}$ by the backward flow, and the point mass eventually
jumps to the neighbouring cell. For $|C'|=10$ this happens at $t\simeq3.3$. The effect requires
an attractor to coincide with a face to within round-off; a shift of the grid by a fraction of a
cell removes it for any number of attractors, and on non-uniform grids each attractor can be
placed at a cell centre. In case B the attractors $\pm\sqrt{1.2}$ do not coincide with
faces of the grids used.}

\begin{table}[t]
\begin{ruledtabular}
\begin{tabular}{lrrlcrr}
run & $N_x$ & ${\rm nnz}/N_x$ & eq. & thr. & \shortstack{ms per\\step} & GB \\
\hline
A, Gaussian & 5200 & 4951 & exact & 2 & 102 & 0.69 \\
A, Gaussian & 5200 & 4951 & ULME & 2 & 94 & 0.69 \\
A, Gaussian & 5200 & 4951 & Poisson & 4 & 45 & 0.69 \\
B, Gaussian & 6000 & 5179 & exact & 2 & 123 & 0.83 \\
B, Laplace & 21783 & 21542 & exact & 6 & 649 & 12.5 \\
B, dich. & 12000 & 3 & exact & 6 & 14 & $<0.01$ \\
near $x^*$, dich. & 42000 & 3 & Poisson & 4 & 1.9 & $<0.01$ \\
\end{tabular}
\end{ruledtabular}
\caption{Cost of representative runs on one node with two Intel Xeon E5-2640 v4 processors:
number of cells $N_x$; total number of non-zero entries (nnz) of the $N_x\times N_x$ kick matrix
divided by $N_x$, i.e. the average number of non-zero entries per row (95\% of all entries for
case A); equation solved; OpenMP threads; elapsed
(wall-clock) time per time step; peak memory. ``dich.'': dichotomous kicks; ``near $x^*$'': the runs of
Sec.~\ref{sec:verification}\,E; ``exact'': Eq.~\eqref{eq:exact} with $M=40$ exponentials. The sparse
products are parallel; transport and auxiliary fields are serial. The peak memory is reached while
the kick matrix is assembled and does not depend on the equation solved; it was measured for the
first and the fifth row and scaled with the number of non-zeros for the fourth.}
\label{tab:cost}
\end{table}

\subsection{The algorithm}
\label{sec:algorithm}

Initialisation: compute $R$ and $R'$ from Eq.~\eqref{eq:volterra}; fit the exponentials;
assemble $K$ from Eq.~\eqref{eq:kernel}; compute the departure points of the faces for a
half step. Each time step then consists of:
\begin{enumerate}
\item drift half step, Eq.~\eqref{eq:csl}, applied to $Q$ and to every $Y_m$;
\item kick step for $Q$ (Heun), with $\Phi$ from Eq.~\eqref{eq:exact}, \eqref{eq:ulme} or
\eqref{eq:poisson} and the first-kick source of Eq.~\eqref{eq:split}, and exact update of the
$Y_m$, driven by $Q$;
\item drift half step as in 1;
\item add the exact never-kicked term when output is required.
\end{enumerate}
The dominant cost is the sparse product with $K$, whose number of non-zeros grows with the
range of the kicks measured in cells (Table~\ref{tab:cost}). For continuous kick distributions
with multiplicative coupling the kernel, truncated at weights $10^{-13}$, is practically dense,
and the cost grows as $N_x^2$; the time per step is then proportional to the number of
non-zeros, about 7--8~ns per non-zero and thread on our hardware for the two products of a Heun
step. The auxiliary fields of the memory add about 30--40~ns per exponential and per cell; with
$M=40$ this is 8\% of the total for Gaussian kicks, and the dominant cost only for discrete kicks,
whose kernel has a few non-zeros per row. The peak memory is about 2.3 times the storage of the
kick matrix, because of the temporary arrays used to assemble it.

Truncating $K$ at larger weights does not reduce the cost appreciably. For case A ($\gamma=1$,
$\mu=2.5$), raising the
threshold from $10^{-13}$ to $10^{-8}$ removes 6\% of the non-zeros, and to $10^{-6}$ only 14\%, while
it already loses $10^{-3}$ of the mass and changes the solution by $W_1/\sigma_P=5\times10^{-2}$,
where $W_1=\int|F-F_{\rm ref}|\,\dd x$ is the Wasserstein distance between the cumulative
distributions of the truncated and the reference solutions, and $\sigma_P$ the standard deviation
of the density: the weights dropped are not redistributed. The kernel is nearly dense not because of the tails of the
kick distribution, but because the domain spans only a few kick widths in the Lamperti variable.
The cost is thus set by the range of the kicks relative to the domain. Two routes reduce it,
for nonlinear as well as linear $I$. Equation~\eqref{eq:kernel} holds for arbitrary cell faces,
and the transport step~\eqref{eq:csl} for non-uniform cells, so that the grid can be refined only
near the attractors, lowering $N_x$ and the cost, which grows as $N_x^2$. On a grid uniform in the
Lamperti variable $G$, moreover, $K_{ij}$ depends only on $i-j$: the kick operator is a discrete
convolution and can be applied by fast Fourier transform in $O(N_x\log N_x)$ operations, for any
$I$ without zeros in the domain.

\emph{Boundaries.} Kicks that would carry mass outside $[x_{\min},x_{\max}]$ remove it from the
domain, and the amount is recorded (absorbing boundary for the kicks); the transport step has zero
flux through both boundary faces. The mass removed would in reality return under a confining drift,
so the boundary lowers the density by a factor that changes slowly with time; the domain must be
large enough for this factor to be negligible for the quantities of interest. Since a kick is a
translation in $G$, the probability that a kick from the bulk $x_b$ leaves the domain is
$1-F_\xi(G(x_{\max})-G(x_b))$; with multiplicative coupling the required $x_{\max}$ grows
exponentially with the number of kick widths, $G=\beta^{-1}\ln(1+\beta x)$ for $I=1+\beta x$.
This estimate ignores sequences of kicks, and the recorded loss is the actual check.
The design and parameter selection of the numerical experiments were assisted by Anthropic's Claude (Opus 5.5 https://claude.ai), under the authors' direction; all simulation setups, code implementations, 
and numerical results were independently executed and verified by the authors.

\section{Verification}
\label{sec:verification}

The method is verified in five steps: against exact solutions of the Poissonian equation and of
the ULME (Sec.~\ref{sec:verification}\,A); against event-driven Monte Carlo simulations with exact
flow maps (B), up to $10^8$ trajectories (C); for kick distributions other than the Gaussian one,
with the numerical parameters fixed beforehand (D); and near a zero of $I$, against an asymptotic
law that extends far below the reach of Monte Carlo sampling (E).

\subsection{Exact solutions}

\begin{figure*}[t]
\centering
\includegraphics[width=\textwidth]{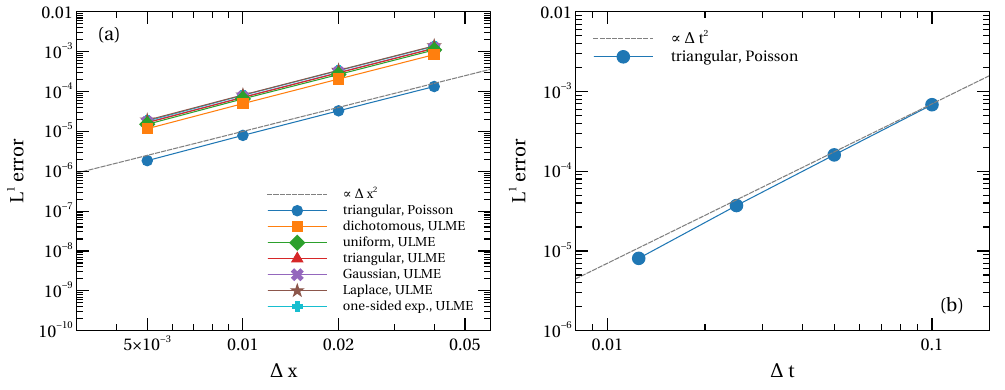}
\caption{$L^1$ error with respect to the exact solution~\eqref{eq:exactk}, $C=0.5x$, $I=1$.
(a) Grid refinement: triangular kicks with Poissonian events ($\langle\tau\rangle=0.5$, $t=10$,
$\Delta t=10^{-3}$), and the six kick distributions of Table~\ref{tab:pdfs} with the ULME (Pareto
WTD, $\mu=1.5$, $T=1$, $t=3$, $\Delta t=2.5\times10^{-3}$). (b) Time-step refinement, triangular
kicks, Poissonian events, $N_x=4000$. Dashed lines: second order.}
\label{fig:convergence}
\end{figure*}

For $C=\gamma x$, $I\equiv I_0$ and $\Phi=r(t)P$ (Poisson or ULME), Eq.~\eqref{eq:me} is solved
by characteristics in Fourier space,
\begin{align}
\tilde P(k,t)&=\tilde P_0\!\left(ke^{-\gamma t}\right)\nonumber\\
&\times\exp\!\left\{\int_0^tr(s)\left[\phat\!\left(I_0ke^{-\gamma(t-s)}\right)-1\right]\dd s\right\},
\label{eq:exactk}
\end{align}
for any kick distribution. On these benchmarks mass is conserved to $10^{-9}$ and the $L^1$ error
converges at second order in $\Delta x$ and in $\Delta t$ (Fig.~\ref{fig:convergence}): each halving
of $\Delta x$ divides the error by 4.0--4.2 for all the distributions of Table~\ref{tab:pdfs},
including dichotomous kicks, which are deposited on the grid with linear weights, and each halving
of $\Delta t$ by 4.3--4.6, for time steps up to $0.2\langle\tau\rangle$.

\subsection{Monte Carlo with exact flow maps}
\label{sec:mc}

When no exact solution is available, the reference is an event-driven simulation of
Eq.~\eqref{eq:sde} without time discretisation of the noise: renewal times are sampled exactly,
the drift between events is its flow (analytic, or adaptive Runge--Kutta with step
$h\le\min(h_0,0.1/|C'|)$), and each kick applies the exact flow map $\phi_{\xi}$; the bias introduced by
the Heun kick commonly used for this purpose is analysed in Ref.~\cite{BianucciMannellaFNL}. The
PDE solver and the simulation read the same input and use the same functions $C$, $I$ and flows.

Let $N$ be the number of simulated trajectories, $P_{\rm MC}$ their histogram, $\sigma_{\rm MC}$
its statistical error in each bin, and $F_{\rm PDE}$, $F_{\rm MC}$ the cumulative distributions of
the PDE solution and of the sample. We use two metrics. The first is the bin-wise residual
$z=(P_{\rm PDE}-P_{\rm MC})/\sigma_{\rm MC}$, whose root mean square over the bins is $\simeq1$ when
the two are statistically indistinguishable. The second is the Kolmogorov distance
$D=\max_x|F_{\rm PDE}-F_{\rm MC}|$, which is robust to quasi-singular components. Sampling noise
alone produces values of $D$ below $b_N=1.36/\sqrt N$ in 95\% of the cases, even when the PDE is
exact (the 95\% band of the Kolmogorov--Smirnov test); we therefore report $D/b_N$, for which
values below one are consistent with an exact PDE.

\subsection{Stress test against $10^8$ trajectories}
\label{sec:stress}

A comparison at a single value of $N$ cannot reveal systematic errors smaller than the statistical
resolution, about $1/\sqrt N$: with $N=10^6$, an error of order $10^{-3}$ goes undetected. We
therefore compared each PDE solution with three independent samples, $N=2\times10^6$,
$2\times10^7$ and $10^8$, and varied one numerical parameter at a time. A systematic error appears as rms $z$ and $D/b_N$ growing as $\sqrt N$; the refinement that stops
the growth identifies its source. We used case A with $\gamma=1$, $\mu=2.5$, $t\le4$, and case B
with $\gamma=1$, $\mu=1.5$, $t\le3$, where half of the mass has never been kicked at $t=3$ and sits
on the attractors $\pm\sqrt{1.2}$.

\begin{table}[t]
\begin{ruledtabular}
\begin{tabular}{lcccccc}
 & \multicolumn{2}{c}{$N=2\times10^6$} & \multicolumn{2}{c}{$N=2\times10^7$} & \multicolumn{2}{c}{$N=10^8$} \\
PDE variant & rms $z$ & $D/b_N$ & rms $z$ & $D/b_N$ & rms $z$ & $D/b_N$ \\
\hline
A base & 1.22 & 1.85 & 1.88 & 5.47 & 3.36 & 12.8 \\
A $\Delta x/2$ & 1.03 & 0.67 & 1.19 & 1.72 & 2.06 & 3.84 \\
A $\Delta x/4$ & 1.03 & 0.67 & 1.15 & 1.72 & 2.00 & 3.84 \\
A $\Delta x/2$, dom & 1.03 & 0.65 & 1.07 & 1.02 & 1.18 & 2.83 \\
A dom, split & 1.02 & 0.65 & 1.04 & 1.21 & 1.10 & 1.66 \\
A $\Delta x/2$, dom, split & 1.02 & 0.65 & 1.02 & 0.66 & 1.05 & 0.55 \\
\hline
B base & 2.59 & 6.67 & 6.76 & 20.7 & 13.7 & 45.5 \\
B $\Delta x/2$ & 1.18 & 1.82 & 2.04 & 5.39 & 3.71 & 11.2 \\
B $\Delta x/4$ & 1.06 & 0.70 & 1.17 & 1.84 & 1.50 & 3.30 \\
B base, split & 1.12 & 1.08 & 1.64 & 3.04 & 2.86 & 5.97 \\
B $\Delta x/2$, split & 1.08 & 0.62 & 1.11 & 1.29 & 1.44 & 2.07 \\
B $\Delta x/4$, split & 1.07 & 0.62 & 1.01 & 0.66 & 1.09 & 0.65 \\
\end{tabular}
\end{ruledtabular}
\caption{Stress test: PDE solutions compared with Monte Carlo samples of $N$ trajectories.
rms $z$: root-mean-square over histogram bins of $z=(P_{\rm PDE}-P_{\rm MC})/\sigma_{\rm MC}$, the
difference in units of the Monte Carlo statistical error (values $\simeq1$: differences
consistent with sampling noise). $D/b_N$: Kolmogorov distance between the cumulative
distributions divided by the 95\% band $b_N=1.36/\sqrt N$ (values $<1$: no detectable
difference). Both are the worst values over the output times. A systematic PDE error makes both
grow as $\sqrt N$ along a row. ``base'': $\Delta x=0.0125$ (A), $0.0061$ (B), $\Delta t=10^{-3}$,
$M=40$; $\Delta t/2$, $M=60$ and a doubled Volterra grid give results identical to ``base'' to
all digits (not shown). $\Delta x/2$, $\Delta x/4$: grid refinement; ``dom'': larger domain;
``split'': never-kicked component treated exactly, Eq.~\eqref{eq:split}.}
\label{tab:stress}
\end{table}

\begin{figure*}[t]
\centering
\includegraphics[width=0.85\textwidth]{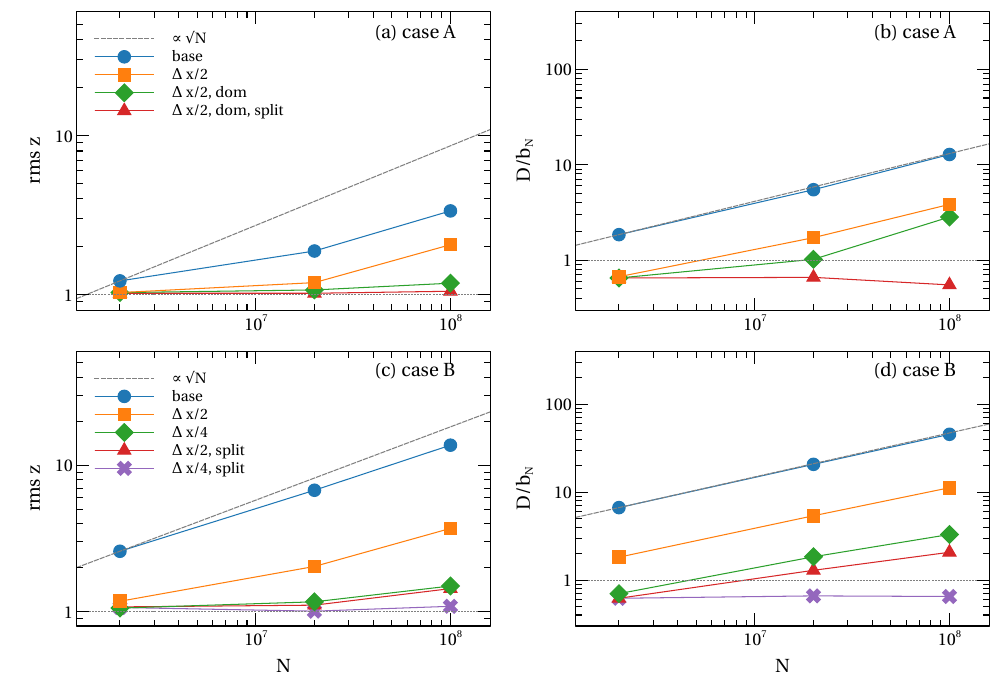}
\caption{Data of Table~\ref{tab:stress} as functions of the number $N$ of Monte Carlo
trajectories: rms $z$ (left) and $D/b_N$ (right) for case A (top) and case B (bottom), along the
refinement ladder. Dotted line: value 1, expected without systematic error. Dashed line:
$\propto\sqrt N$, the growth produced by a systematic error.}
\label{fig:stress}
\end{figure*}

Four conclusions follow from Table~\ref{tab:stress}. (i) Time step, number of exponentials and Volterra
grid are irrelevant even at $N=10^8$. (ii) All systematic error is spatial and associated with
quasi-singular structures: in case B, enlarging the domain at fixed $\Delta x$ makes the error
\emph{worse} (rms $z$ from 13.7 to 21.3 at $N=10^8$, run not shown in the table), because it shifts the cells relative to the attractor.
(iii) The exact never-kicked term removes the dominant part; the remainder comes from kicked mass
re-contracted toward the attractors during long waiting times, and it disappears on refinement.
(iv) With the split and adequate resolution, the ratio $D/b_N$ remains below one and
does not grow with $N$ (0.62, 0.66 and 0.65 for $N=2\times10^6$, $2\times10^7$ and $10^8$ in
case B). A systematic error $\epsilon$ would make this ratio grow as $\epsilon\sqrt N$; its absence
bounds the systematic error of the PDE by the statistical resolution of $10^8$ trajectories,
$\epsilon\lesssim1.4\times10^{-4}$. The residual rms $z\simeq1.05$, which does not grow with $N$
either, reflects the bin-wise variance estimate. The Monte Carlo was itself checked by
replacing the exact flows with an embedded Cash--Karp integrator, without detectable change.
The same data are shown in Fig.~\ref{fig:stress}.

\subsection{Other kick distributions}

The numerical parameters of Sec.~\ref{sec:stress} were determined for Gaussian kicks. To check
that they carry over, we solved case B with the same grid, time step and number of exponentials for
the other distributions of Table~\ref{tab:pdfs}, including the one-sided exponential with
$\xi<0$, which pushes the trajectories toward the zero of $I$, and compared with Monte Carlo
samples of $N=2\times10^7$ trajectories. Only the domain was adapted, for Laplace kicks, to
contain the upper tail produced by the multiplicative coupling. The results are in
Table~\ref{tab:kicks}: over all distributions and output times, rms $z$ lies between 0.91 and 1.07
and $D/b_N$ does not exceed 1.03. With a 95\% band, about one comparison in twenty is expected
above one; among 24 comparisons one reaches 1.03. The parameters determined for Gaussian kicks are
therefore adequate for all the distributions considered, including the asymmetric one.

\begin{figure*}[t]
\centering
\includegraphics[width=0.85\textwidth]{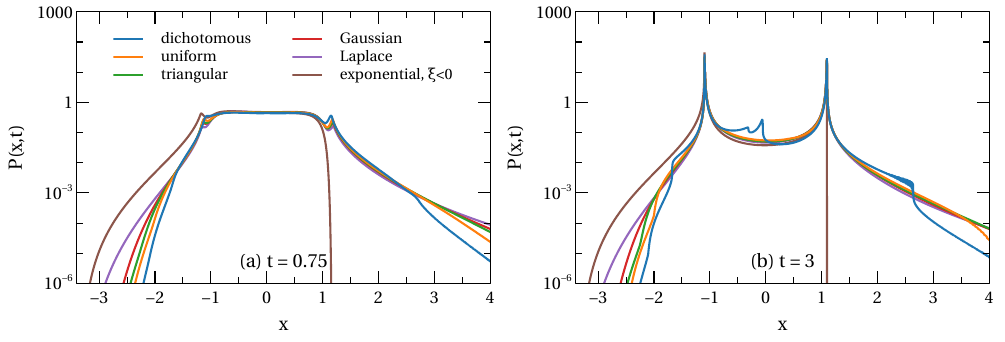}
\caption{Case B ($C=-1.2x+x^3$, $I=1+0.3x$, Pareto WTD with $\mu=1.5$), exact equation, for the kick
distributions of Table~\ref{tab:pdfs}, all of unit standard deviation, at $t=0.75$ and $t=3$.
The secondary peaks of the dichotomous solution, at $x\simeq-0.05$ and $x\simeq2.65$ (images of
the attractor $\sqrt{1.2}$ under kicks $\mp1$), contain never-kicked mass kicked once and not yet
relaxed onto the attractors.}
\label{fig:kicks}
\end{figure*}

The densities themselves (Fig.~\ref{fig:kicks}) show how the statistics of the kicks beyond their
variance shapes the solution. Among the symmetric distributions, the occupation of the outer tails
decreases monotonically with the kurtosis: at $t=3$ the probability of $x>1.5$ goes from
$8.1\times10^{-3}$ for dichotomous kicks (kurtosis 1) to $5.2\times10^{-3}$ for Laplace kicks
(kurtosis 6), and that of $x<-1.5$ from $3.0\times10^{-3}$ to $1.9\times10^{-3}$. The Wasserstein
distance from the solution with Gaussian kicks is smallest for triangular kicks, whose kurtosis
is closest to the Gaussian one ($5.7\times10^{-4}$, against $6.6\times10^{-3}$ for uniform,
$6.9\times10^{-3}$ for Laplace and $4.3\times10^{-2}$ for dichotomous kicks). One-sided kicks with
$\xi<0$ break the symmetry of the drift: the mean is $-0.42$ at $t=3$, and no mass appears beyond
the right attractor.

\begin{table}[t]
\begin{ruledtabular}
\begin{tabular}{lcc}
kick distribution & rms $z$ & $D/b_N$ \\
\hline
dichotomous & 1.07 & 0.56 \\
uniform & 1.07 & 0.59 \\
triangular & 1.06 & 1.03 \\
Gaussian & 1.01 & 0.66 \\
Laplace & 1.03 & 0.78 \\
one-sided exponential, $\xi<0$ & 1.03 & 0.86 \\
\end{tabular}
\end{ruledtabular}
\caption{Case B ($C=-1.2x+x^3$, $I=1+0.3x$, Pareto WTD with $\mu=1.5$, $t\le3$) for different kick
distributions of unit standard deviation, solved with the numerical parameters fixed for Gaussian
kicks in Sec.~\ref{sec:stress}, against Monte Carlo samples of $N=2\times10^7$ trajectories.
Worst values over the output times $t=0.75,1.5,2.25,3$.}
\label{tab:kicks}
\end{table}

\subsection{Near a zero of $I$}

\begin{figure}[t]
\centering
\includegraphics[width=\columnwidth]{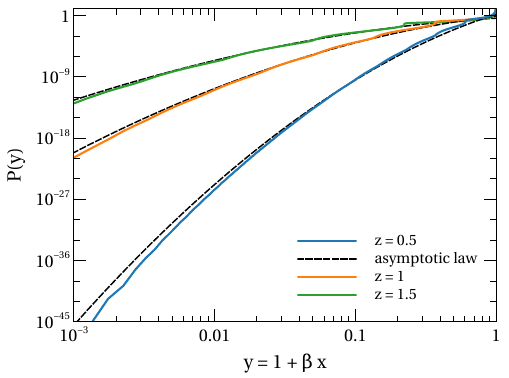}
\caption{Stationary density near the zero $x^*=-2$ of $I=1+0.5x$, as a function of $y=1+\beta x$,
for $C=x$, exponential waiting times with $\gamma\tau=4$ and dichotomous kicks with $z=a\beta=0.5$, 1,
1.5. Solid lines: PDE. Dashed lines: asymptotic law~\eqref{eq:barrier} of
Ref.~\cite{BianucciMannellaFNL}, with fitted amplitude.}
\label{fig:barrier}
\end{figure}

A zero $x^*$ of $I$ is a barrier for the kicks \cite{BianucciMannellaFNL}, and with a drift
pointing away from it the density vanishes there faster than any power. For $C=\gamma x$,
$I=1+\beta x$, dichotomous kicks of amplitude $a$ and exponential waiting times of mean $\tau$,
Ref.~\cite{BianucciMannellaFNL} gives the asymptotic stationary law as $y=1+\beta x\to0$,
\begin{align}
P(y)\simeq A\exp\!\Big\{-\frac{1}{z}\Big[&\frac{\ell^2}{2}+c\,\ell\nonumber\\
&+(\ell+L)\Big(\ln\frac{\ell+L}{z}-1\Big)\Big]\Big\},
\label{eq:barrier}
\end{align}
with $\ell=\ln(1/y)$, $z=a\beta$, $L=\ln(2\gamma\tau)$, $c=L-z/2$. This regime is out of reach of
Monte Carlo simulations, but not of the PDE. Figure~\ref{fig:barrier} shows the stationary
density for $\beta=0.5$, $\gamma\tau=4$ and $z=0.5$, 1 and 1.5, computed with $\Delta x=10^{-3}$,
$\Delta t=10^{-3}$ up to $t=128$, together with Eq.~\eqref{eq:barrier}, whose amplitude $A$ is
fitted for $\ell\ge z+1$. The PDE follows the asymptotic law down to $P\simeq5\times10^{-38}$
for $z=0.5$; the residual in $\ln P$ drifts slowly, reaching 3.5 at $y=3\times10^{-3}$ where
$\ln P\simeq-86$, as expected for an expansion that neglects subleading terms in $\ell$. For
$z=1$ and $1.5$ the residual stays below 1. For $z=1.5$ the second moment of the stationary
density is infinite, because $\gamma\tau=4$ is below $\frac{1}{2}(\cosh2z-1)\simeq4.5$, and the upper
tail keeps spreading; by $t=128$ about 29\% of the mass has left the domain through its upper
boundary at $x=40$ (1\% for $z=1$). Near the barrier this only lowers the amplitude: between
$t=96$ and $t=128$, $\ln P$ decreases uniformly by $0.086$ across the fitting window, with a
spread of $10^{-8}$, so that the shape of the density, which is what Eq.~\eqref{eq:barrier}
describes, is stationary. Quantities sensitive to the upper tail, such as the moments, would
instead depend on $x_{\max}$; for $z=1.5$ the second moment diverges and no finite domain would
make it converge.

Near such a barrier the splitting has a visible effect. The drift moves the density away from
$x^*$ with speed $|C(x^*)|$ and nothing flows in from the other side, so each drift half step
empties a layer of width $|C(x^*)|\Delta t/2$ next to the barrier. The density is reliable only
at distances from $x^*$ larger than a few times this width, which sets the time step required
to resolve the approach to the barrier.

\begin{table}[t]
\begin{ruledtabular}
\begin{tabular}{lcc}
distribution & support / scale & kurtosis \\
\hline
dichotomous & $\pm1$ & 1 \\
uniform & $[-\sqrt3,\sqrt3]$ & 1.8 \\
triangular & $[-\sqrt6,\sqrt6]$ & 2.4 \\
Gaussian & $s=1$ & 3 \\
Laplace & $b=1/\sqrt2$ & 6 \\
one-sided exponential & mean $=1$ & 9 (skewness 2) \\
\end{tabular}
\end{ruledtabular}
\caption{Kick distributions at unit variance. Characteristic functions and cumulative
distributions are in Appendix~\ref{app:kernel}.}
\label{tab:pdfs}
\end{table}


\section{Application: exact, local and Poissonian equations}
\label{sec:examples}

The solver allows a direct comparison of Eq.~\eqref{eq:exact} with its approximations,
Eqs.~\eqref{eq:ulme} and~\eqref{eq:poisson}. Part of the difference is analytic: the never-kicked weight is
$\Psi(t)=(1+t/T)^{1-\mu}$ in the exact equation, but $\exp[-\int_0^tR(s)\,\dd s]$ in the ULME, a stretched
exponential $\exp(-ct^{\mu-1})$ for $\mu<2$ and an exponential for $\mu>2$. The ULME misses the
power-law tail of the deterministic component; we denote the difference of the two weights by
$\Delta S(t)=\Psi(t)-\exp[-\int_0^tR(s)\,\dd s]$. Table~\ref{tab:ulme} gives the distance between
the two equations for case A, varying $\gamma$ and $\mu$.

\begin{table}[t]
\begin{ruledtabular}
\begin{tabular}{lcccc}
$\mu$ & $\gamma T=0.3$ & $1$ & $3$ & $10$ \\
\hline
1.3 & 0.042 & 0.021 & 0.0065 & (0.0015) \\
1.5 & 0.063 & 0.038 & 0.013 & 0.0032 \\
1.8 & 0.076 & 0.060 & 0.024 & 0.0061 \\
2.2 & 0.066 & 0.075 & 0.036 & 0.010 \\
2.5 & 0.061 & 0.078 & 0.043 & 0.013 \\
3.0 & 0.053 & 0.077 & 0.048 & 0.017 \\
\end{tabular}
\end{ruledtabular}
\caption{Wasserstein distance between exact equation and ULME in units of the standard deviation
of the exact solution, $W_1/\sigma_P$, maximum over $t\le8$; case A with $C=\gamma x$, $I=1+0.5x$,
Gaussian kicks, $T=1$. The numerical floor, measured with nested grids refined $\times3$, is
below 10\% of the entries, except for $\gamma T=10$, $\mu=1.3$ (in parentheses), which is at the
floor. Column $\gamma T=0.3$, $\mu\ge2.2$: domain extended to $x\simeq150$.}
\label{tab:ulme}
\end{table}

The choice of the distance matters here. Both densities contain near point masses at the
attractor, where the cumulative distributions jump; the Kolmogorov distance, a supremum over
$x$, then depends on where the grid samples the jump, and in our runs it changed by 14--42\%
under a nested $\times3$ refinement even for $\gamma T=3$. The Wasserstein distance
$W_1=\int|F_{\rm exact}-F_{\rm ULME}|\,\dd x$ is insensitive to the redistribution of a point
mass within a cell, and it is resolved in all cases of Table~\ref{tab:ulme} except one.

The local equation differs from the exact one for all $\mu$, including $\mu>2$ where the mean
waiting time is finite, and well above the numerical floor. The wrong never-kicked weight
matters only when the unkicked mass can be told apart from the kicked one. For $\gamma T=10$
the kicked mass relaxes onto the attractor $x=0$ long before the next kick: for $\mu=1.3$ the
two equations are indistinguishable at our resolution, although their never-kicked weights
differ by $\Delta S\simeq0.06$ at $t=8$. This case also shows that the resolution required near an
attractor grows with the contraction rate $|C'(x^*)|$.

In units of the width of the density (Table~\ref{tab:ulme}), the deviation decreases with $\gamma T$:
when the drift relaxes the system between kicks, the transport during the waiting times no longer
matters, and at $\gamma T=10$ the deviation is about $1\%$ of $\sigma_P$ or less. For $\gamma T\ge1$ it
grows with $\mu$, up to $7$--$8\%$ of $\sigma_P$, and it remains of this size for $\mu>2$, where the
mean waiting time is finite; for $\mu\ge2.2$ it is largest at $\gamma T\simeq1$. With the bistable
drift $C=\gamma(-1.2x+x^3)$, $I=1+0.3x$, the deviation is $0.9$--$1.5\%$ of $\sigma_P$ for $\mu=1.5$
and $3$--$4.6\%$ for $\mu=2.5$, again decreasing with $\gamma$.

For $\mu>2$ the deviation depends essentially on one ratio, the drift relaxation time
$1/|C'(x^*)|=1/\gamma$ against the scale $T$ of the waiting times: $W_1/\sigma_P=0.075$--$0.078$ at
$\gamma T=1$, $0.036$--$0.048$ at $\gamma T=3$ and $0.010$--$0.017$ at $\gamma T=10$ for $\mu=2.2$,
2.5 and 3, while the mean waiting time $\langle\tau\rangle=T/(\mu-2)$ changes by a factor of five.
The relevant time is thus that of the short waiting times, not the mean; in case A a deviation
below about $2\%$ of $\sigma_P$ requires $|C'(x^*)|T\gtrsim10$. In the bistable case B, with
$|C'(x^*)|=2.4\gamma$, the deviations are of the same order but do not fall on the same curve,
so that this criterion is indicative.

These trends agree with an estimate of the error of the ULME \cite{Bianucci2026corr}. The
exact kick flux equals the ULME flux plus a remainder of second order in $\phat-1$, whose
relative size is $\varepsilon\sim\|\phat-1\|\,\Delta$, with
$\Delta=\int_0^{\tau_C}[R(v)-R(t)]\,\dd v$ the excess number of renewals within the relaxation
time $\tau_C\simeq1/|C'(x^*)|$ that follows a renewal. For $\tau_C\ll T$,
$\Delta\simeq\tau_C[R(0)-R(t)]$. For the Pareto WTD $R(0)=(\mu-1)/T$, so that for $\mu>2$
$R(0)-1/\langle\tau\rangle=1/T$ independently of $\mu$: the relevant scale is $T$, not
$\langle\tau\rangle$, as found above. For $\mu<2$, $R(t)\to0$ and $\Delta\simeq(\mu-1)\tau_C/T$,
which accounts qualitatively for the increase with $\mu$ of the column $\gamma T=10$ of
Table~\ref{tab:ulme} and for its slower growth for $\mu>2$. The same estimate predicts larger
deviations in the opposite regime, $T\ll\tau_C\ll t$, especially for $\mu<2$, where $\Delta$
grows as $(\tau_C/T)^{\mu-1}$; this regime lies outside the range of Table~\ref{tab:ulme}.
Finally, $W_1/\sigma_P$ measures the displacement of the bulk of the density; the tails can
differ more: for $\mu=1.5$, $\gamma T=1$ and the linear drift, the ULME underestimates the density
at $x=8$ by a factor of about 2 at $t=8$, against 1.4 at $t=2$ (Fig.~S1 of the Supplemental
Material~\cite{SM}).

Figure~\ref{fig:densities} shows the densities of the three equations for $\mu=2.5$ and $\gamma T=1$,
with the linear drift of case A and the bistable drift of case B. At $t=2$ the ULME, which
contains the actual renewal rate $R(t)$, is much closer to the exact density than the Poissonian
equation, which replaces it with $1/\langle\tau\rangle$: the Wasserstein distances from the exact
solution are $0.043$ and $0.195$ in case A, $0.009$ and $0.044$ in case B. At $t=8$, as $R(t)$
approaches $1/\langle\tau\rangle$, the two approximations become comparable ($0.081$ and $0.098$ in
case A for the ULME and the Poissonian equation), and in case B the Poissonian equation is the
closer of the two ($0.011$, against $0.041$ for the ULME). Neither contains the transport during the waiting times, and which of the two errors is
smaller depends on the case.

For $\mu\le2$ the mean waiting time is infinite and only the comparison with the ULME is possible
($\mu=1.5$; the densities are shown in the Supplemental Material~\cite{SM}). Here the never-kicked mass is large, and its weight differs
markedly between the two equations: $\Psi(8)=0.33$ in the exact one, $\exp[-\int_0^8R(s)\,\dd s]=0.23$
in the ULME. The distance between the two solutions grows with time, from $W_1/\sigma_P=0.020$ at
$t=2$ to $0.038$ at $t=8$ with the linear drift, and from $0.002$ to $0.011$ with the bistable one.

\begin{figure*}[t]
\centering
\includegraphics[width=0.85\textwidth]{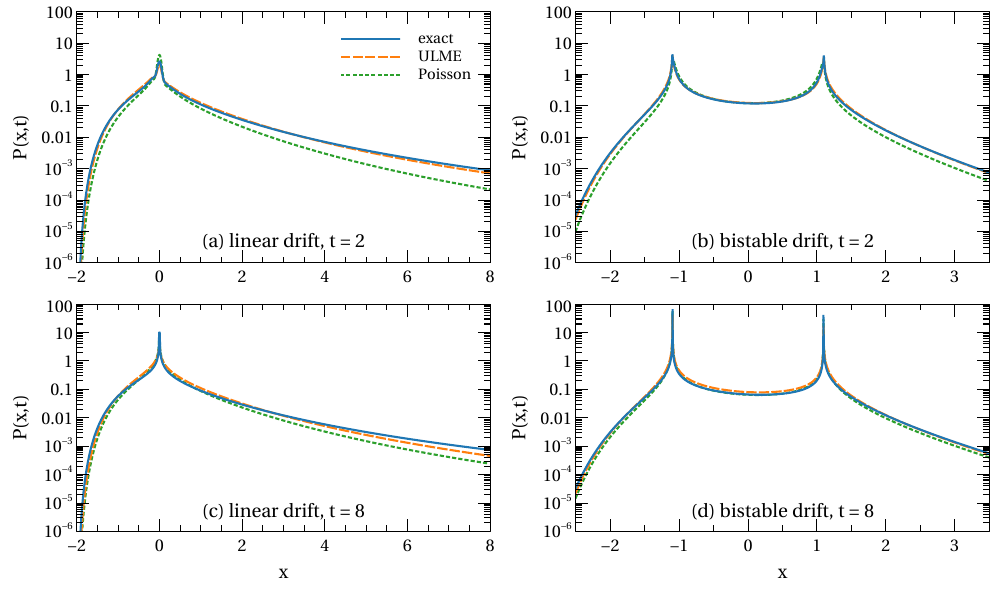}
\caption{Densities of the exact equation~\eqref{eq:exact} (solid), the ULME~\eqref{eq:ulme}
(dashed) and the Poissonian equation~\eqref{eq:poisson} (dotted), for $\mu=2.5$, $T=1$, $\gamma=1$,
Gaussian kicks. (a), (c): $C=x$, $I=1+0.5x$; (b), (d): $C=-1.2x+x^3$, $I=1+0.3x$; at $t=2$ (top)
and $t=8$ (bottom). The peaks contain the never-kicked mass, collapsed onto the attractors.}
\label{fig:densities}
\end{figure*}

\section{Numerical parameters}
\label{sec:recommendations}

In all cases studied the accuracy was controlled by the spatial grid near the attractors of the
drift, not by the time step: $\Delta x$ between $1.5\times10^{-3}$ and $6\times10^{-3}$ was
required near the fixed points, with the never-kicked mass treated exactly and no attractor
on a cell face. The time step and the number of exponentials could be kept at moderate values
($\Delta t=10^{-3}T$, $M=40$); halving $\Delta t$ or increasing $M$ to 60 had no detectable
effect. With multiplicative kicks the upper tail of the density can extend far beyond the bulk,
and the domain was chosen so that the mass leaving it remained well below the quantities of
interest. The cell-integrated kernel required no tuning, whereas Gauss rules in $\xi$ required
a number of nodes dependent on the structure of the solution. Comparison with Monte Carlo
simulations at increasing $N$ was the most sensitive test: a systematic error shows up as a
discrepancy growing as $\sqrt N$.

\section{Conclusions}
\label{sec:conclusions}

We have described a method for the numerical integration of the master equations of
Ref.~\cite{Bianucci2026}, which govern systems driven by a drift and by renewal trains of
state-dependent kicks. Its ingredients are a cell-integrated kick operator obtained from the
Lamperti variable, a conservative semi-Lagrangian transport exact in time, a sum-of-exponentials
memory whose auxiliary fields are carried by the drift, and the exact treatment of the
never-kicked mass. The resulting scheme is second order in space and time on exact benchmarks,
handles nonlinear drifts and couplings and arbitrary kick distributions, and costs, per time step,
a sparse product with the kick matrix and a fixed number of transport steps, independently of the
length of the history.

The verification has shown where the numerical difficulty lies. Time step, memory representation
and renewal density contributed no detectable error even against $10^8$ trajectories. The error
was spatial, and it came from quasi-singular structures: the never-kicked mass collapsing onto the
attractors of the drift and, to a lesser extent, kicked mass re-contracted toward them during long
waiting times. The same structures affect the measures used to compare solutions: distances based
on the supremum of the difference between cumulative distributions depend on how a near point mass
is sampled by the grid, while the Wasserstein distance does not. Near a zero of the coupling with
the drift pointing away from it, the splitting empties a layer of width $|C(x^*)|\Delta t/2$, which
sets the time step needed to follow the density toward the barrier.

With these choices the method reproduces event-driven simulations with exact flow maps to within
their statistical resolution and extends far beyond it: near a zero of $I$ it follows the
asymptotic stationary law over more than thirty decades. It also allows a direct assessment of
approximate descriptions. The universal local master equation, exact for Poissonian events
and to first order in the kick operator, deviates from the exact equation, in the range
explored, by up to $8\%$ of the width of the density when the relaxation time of the drift is
comparable with the short waiting times, also when their mean is finite; the deviation becomes
negligible when the drift relaxes the system between kicks. An estimate of its error predicts
larger deviations for drifts much slower than the scale $T$ of the waiting times, at long
times.

Several extensions are straightforward. Non-uniform grids, refined near the attractors, would
reduce the cost of resolving the quasi-singular structures; on a grid uniform in the Lamperti
variable the kick matrix is a convolution and can be applied by fast Fourier transform, for any
$I$ without zeros. Kick distributions with heavy tails, for which the Kramers--Moyal expansion of
$\phat(i\pa_xI)$ does not exist, require no change, since only the cumulative distribution enters
Eq.~\eqref{eq:kernel}. Since Eqs.~\eqref{eq:flux} require only a propagator between kicks, further
cases have the same structure, with the auxiliary fields transported by the corresponding
operator: an additional Markovian background noise, time-dependent drifts, absorbing boundaries
and reset, and, in several dimensions, kicks acting along one coordinate. These extensions will
be the subject of future work.

\begin{acknowledgments}
We thank the Green Data Center of the University of Pisa for computational resources. During the preparation 
of this manuscript, the authors used Anthropic's Claude (Opus 5.5, https://claude.ai) for structural feedback 
and critical revision; the authors reviewed, edited, and verified all text and take 
full responsibility for the content of the publication.
\end{acknowledgments}

\appendix

\section{Kick distributions and the cell-integrated kernel}
\label{app:kernel}

For a kick distribution of standard deviation $s$: uniform on $[-a,a]$, $a=\sqrt3s$,
$F(\xi)=(\xi+a)/(2a)$, $\phat(k)=\sin(ak)/(ak)$; triangular, $p(\xi)=(a-|\xi|)/a^2$, $a=\sqrt6s$,
$\phat(k)=[\sin(ak/2)/(ak/2)]^2$; Gaussian, $\phat(k)=e^{-s^2k^2/2}$; Laplace,
$p(\xi)=e^{-|\xi|/b}/(2b)$, $b=s/\sqrt2$, $\phat(k)=1/(1+b^2k^2)$; one-sided exponential with mean
$s$, $\phat(k)=1/(1-isk)$. In Eq.~\eqref{eq:kernel}, differences of $F$ in the upper tail are
evaluated through the survival function to avoid cancellation.

Table~\ref{tab:kernel} compares the cell-integrated kernel with Gauss rules in $\xi$ (Hermite for
Gaussian kicks, Laguerre on each half-line for the others) on the exact ULME benchmark of
Fig.~\ref{fig:convergence}. The cell-integrated kernel gives the smallest error, which is the
spatial discretisation error at this resolution; the Gauss rules approach it as the number of
nodes grows, at a rate that depends on the distribution.

\begin{table}[t]
\begin{ruledtabular}
\begin{tabular}{lccccc}
 & \multicolumn{4}{c}{Gauss rule, nodes} & \\
kicks & 30 & 60 & 120 & 240 & cell \\
\hline
Gaussian & 124 & 8.10 & 7.37 & 7.41 & 7.24 \\
Laplace & 38.2 & 10.8 & 8.56 & 8.47 & 8.19 \\
one-sided exp. & 129 & 27.5 & 9.95 & 8.22 & 8.05 \\
\end{tabular}
\end{ruledtabular}
\caption{$L^1$ error, in units of $10^{-5}$, at $t=3$ on the exact ULME benchmark ($C=0.5x$, $I=1$, Pareto WTD with
$\mu=1.5$, $N_x=4000$), for Gauss rules in $\xi$ with increasing number of nodes and for the
cell-integrated kernel~\eqref{eq:kernel}.}
\label{tab:kernel}
\end{table}

\section{Auxiliary fields: exponential integrator and fit of the kernel}
\label{app:soe}

Over a step $h$, with $P$ linear in time between $P^n$ and $P^{n+1}$,
$Y_m^{n+1}=e^{-\lambda_mh}Y_m^n+c_m\left(a_1P^n+a_2P^{n+1}\right)$,
$a_2=(\lambda_mh-1+e^{-\lambda_mh})/(\lambda_m^2h)$, $a_1=(1-e^{-\lambda_mh})/\lambda_m-a_2$, stable
for any $\lambda_mh$.

The rates $\lambda_m$ are log-spaced in $[0.05/t_{\max},\min(4/\Delta t,10^3/T)]$. The amplitudes
$c_m$ minimise
\begin{equation}
\sum_{k=1}^{n_s}w_k^2\Big[\sum_{m=1}^Mc_me^{-\lambda_mu_k}-R'(u_k)\Big]^2
+\varepsilon^2\sum_{m=1}^M\|\mathbf v_m\|^2c_m^2
\end{equation}
over $n_s=25M$ points $u_k$, with $u_1=0$ and the others log-spaced up to $t_{\max}$. The weights
$w_k=1/\max\big(|R'(u_k)|,10^{-6}\max_j|R'(u_j)|\big)$ make the fit relative, so that the slow tail
of $R'$ is fitted as accurately as its short-time part; the floor avoids divergent weights where
$R'$ vanishes. The Tikhonov term, with $\varepsilon=10^{-8}$ and $\mathbf v_m$ the vector of
components $w_ke^{-\lambda_mu_k}$, i.e.\ the $m$th column of the weighted system, prevents large
amplitudes of alternating sign, to which the nearly collinear exponentials are prone. The system
is solved by Householder QR factorisation; the code prints the maximum relative error of the fit,
which was $\simeq3\times10^{-9}$ for $M=40$.

\section{Code}

The F90 solver, the Monte Carlo reference, the input files of all runs, the analysis
scripts and the data of the figures are available on Zenodo, \href{https://doi.org/10.5281/zenodo.23856371}{doi:10.5281/zenodo.23856371}. The code uses the same sign convention as the text,
$\dot x=-C+I\xi$.

\section{Supplemental Material}
\begin{figure*}[htb]
\centering
\includegraphics[width=0.85\textwidth]{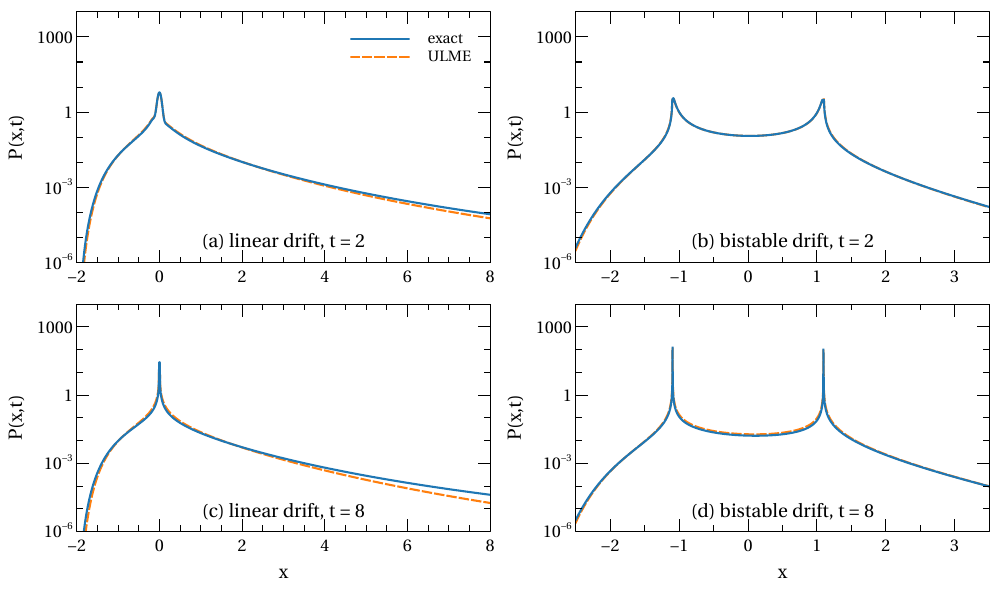}
\caption{Densities of the exact equation (solid) and of the ULME (dashed) for a Pareto waiting-time
distribution with $\mu=1.5$, $T=1$, $\gamma=1$, Gaussian kicks of unit variance. (a), (c): $C=x$,
$I=1+0.5x$; (b), (d): $C=-1.2x+x^3$, $I=1+0.3x$; at $t=2$ (top) and $t=8$ (bottom).}
\label{fig:S1}
\end{figure*}

Figure~\ref{fig:S1} complements Sec.~V of the main text for a waiting-time distribution with infinite
mean, $\mu=1.5$, for which the Poissonian equation does not exist. The never-kicked mass, collected in
the peaks at the attractors, has weight $\Psi(8)=0.33$ in the exact equation and
$\exp[-\int_0^8R(s)\,\mathrm{d}s]=0.23$ in the universal local master equation (ULME). The Wasserstein
distance between the two solutions, in units of the standard deviation of the exact one, grows from
0.020 at $t=2$ to 0.038 at $t=8$ with the linear drift, and from 0.002 to 0.011 with the bistable drift.

\bibliography{refs_}

@article{Bianucci2026,
  author  = {Bianucci, M. and Bologna, M. and Mannella, R.},
  title   = {Exact and limit results for the {CTRW} in presence of drift and position dependent noise intensity},
  journal = {J. Stat. Mech.},
  year    = {2026},
  pages   = {073203},
  doi     = {10.1088/1742-5468/ae7bd9}
}

@unpublished{Bianucci2026corr,
  author  = {Bianucci, M. and Bologna, M. and Mannella, R.},
  title   = {Corrigendum: Exact and limit results for the {CTRW} in presence of drift and position dependent noise intensity},
  note    = {submitted to J. Stat. Mech.},
  year    = {2026}
}

@unpublished{BianucciMannellaFNL,
  author  = {Bianucci, M. and Mannella, R.},
  title   = {It\^o versus {S}tratonovich for shot noise with state-dependent amplitude: why the answer is {M}arcus},
  note    = {submitted to Fluct. Noise Lett.},
  year    = {2026}
}

@article{DiPaolaFalsone1993,
  author  = {Di Paola, M. and Falsone, G.},
  title   = {It\^o and {S}tratonovich integrals for delta-correlated processes},
  journal = {Probab. Eng. Mech.},
  volume  = {8},
  pages   = {197},
  year    = {1993},
 DOI = {10.1016/0266-8920(93)90015-N}
}

@article{Marcus1978,
  author  = {Marcus, S. I.},
  title   = {Modeling and analysis of stochastic differential equations driven by point processes},
  journal = {IEEE Trans. Inf. Theory},
  volume  = {24},
  pages   = {164},
  year    = {1978},
DOI = {10.1109/TIT.1978.1055857}
}

@article{Marcus1981,
  author  = {Marcus, S. I.},
  title   = {Modeling and approximation of stochastic differential equations driven by semimartingales},
  journal = {Stochastics},
  volume  = {4},
  pages   = {223},
  year    = {1981},
 DOI = {10.1080/17442508108833165}
}

@article{ChechkinPavlyukevich2014,
  author  = {Chechkin, A. V. and Pavlyukevich, I.},
  title   = {Marcus versus {S}tratonovich for systems with jump noise},
  journal = {J. Phys. A: Math. Theor.},
  volume  = {47},
  pages   = {342001},
  year    = {2014},
 DOI = {10.1088/1751-8113/47/34/342001}
}

@article{DenisovHorsthemkeHanggi2009,
  author  = {Denisov, S. I. and Horsthemke, W. and H{\"a}nggi, P.},
  title   = {Generalized {F}okker-{P}lanck equation: Derivation and exact solutions},
  journal = {Eur. Phys. J. B},
  volume  = {68},
  pages   = {567},
  year    = {2009},
  DOI = {10.1140/epjb/e2009-00126-3}
}

@article{SunLiZheng2014,
  author  = {Sun, Xu and Duan, Jinqiao and Li, Xiaofan and Liu, Hua and Wang, Xiangjun and Zheng, Yayun},
  title   = {Derivation of {Fokker--Planck} equations for stochastic systems under excitation of multiplicative non-{Gaussian} white noise},
  journal = {J. Math. Anal. Appl.},
  volume  = {446},
  number  = {1},
  pages   = {786--800},
  year    = {2017},
  doi     = {10.1016/j.jmaa.2016.09.010}
}

@article{DiMatteoDiPaolaPirrotta,
  author  = {Di Matteo, A. and Di Paola, M. and Pirrotta, A.},
  title   = {Path integral solution for nonlinear systems under parametric {P}oissonian white noise input},
  volume = {44},
  pages   = {89--98},
  journal = {Probab. Eng. Mech.},
  year    = {2016},
  doi     = {10.1016/j.probengmech.2015.09.020}
}

@article{LaiDeKamps2016,
  author  = {Lai, Y. M. and de Kamps, M.},
  title   = {Population density equations for stochastic processes with memory kernels},
  journal = {Phys. Rev. E},
  volume = {95},
  number = {6},
  pages = {062125},
  numpages = {11},
  year = {2017},
  month = {Jun},
  publisher = {American Physical Society},
  doi = {10.1103/PhysRevE.95.062125},
  url = {https://link.aps.org/doi/10.1103/PhysRevE.95.062125}
}

@article{MontrollWeiss1965,
  author  = {Montroll, E. W. and Weiss, G. H.},
  title   = {Random walks on lattices. {II}},
  journal = {J. Math. Phys.},
  volume  = {6},
  pages   = {167},
  year    = {1965},
  DOI = {10.1063/1.1704269}
}

@article{MetzlerBarkaiKlafter1999,
  author  = {Metzler, R. and Barkai, E. and Klafter, J.},
  title   = {Anomalous diffusion and relaxation close to thermal equilibrium: A fractional {F}okker-{P}lanck equation approach},
  journal = {Phys. Rev. Lett.},
  volume  = {82},
  pages   = {3563},
  year    = {1999},
  DOI = {10.1103/PhysRevLett.82.3563}
}

@article{SokolovKlafter2006,
  author  = {Sokolov, I. M. and Klafter, J.},
  title   = {Field-induced dispersion in subdiffusion},
  journal = {Phys. Rev. Lett.},
  volume  = {97},
  pages   = {140602},
  year    = {2006}
}

@article{Heinsalu2007,
  author  = {Heinsalu, E. and Patriarca, M. and Goychuk, I. and H{\"a}nggi, P.},
  title   = {Use and abuse of a fractional {F}okker-{P}lanck dynamics for time-dependent driving},
  journal = {Phys. Rev. Lett.},
  volume  = {99},
  pages   = {120602},
  year    = {2007}
}

@article{Crouseilles2010,
  author  = {Crouseilles, N. and Mehrenberger, M. and Sonnendr{\"u}cker, E.},
  title   = {Conservative semi-{L}agrangian schemes for {V}lasov equations},
  journal = {J. Comput. Phys.},
  volume  = {229},
  pages   = {1927},
  year    = {2010},
  DOI = {10.1016/j.jcp.2009.11.007}
}

@article{Lubich1988,
  author  = {Lubich, C.},
  title   = {Convolution quadrature and discretized operational calculus. {I}},
  journal = {Numer. Math.},
  volume  = {52},
  pages   = {129},
  DOI = {10.1007/BF01398686},
  year    = {1988}
}

@article{Jiang2017,
  author  = {Jiang, S. and Zhang, J. and Zhang, Q. and Zhang, Z.},
  title   = {Fast evaluation of the {C}aputo fractional derivative and its applications to fractional diffusion equations},
  volume={21}, 
  DOI={10.4208/cicp.OA-2016-0136}, 
  number={3}, 
  year={2017}, 
  pages={650--678},
  journal = {Commun. Comput. Phys.},
}

@article{dHalluin2005,
  author  = {d'Halluin, Y. and Forsyth, P. A. and Vetzal, K. R.},
  title   = {Robust numerical methods for contingent claims under jump diffusion processes},
  journal = {IMA J. Numer. Anal.},
  volume  = {25},
  number  = {1},
  pages   = {87--112},
  year    = {2005},
  doi     = {10.1093/imanum/drh011}
}

@book{Cox1962,
  author    = {Cox, D. R.},
  title     = {Renewal Theory},
  publisher = {Methuen},
  address   = {London},
  year      = {1962}
}

@misc{SM,
  note = {See Supplemental Material at [URL will be inserted by publisher] for the densities of the exact equation and of the ULME for $\mu=1.5$}
}

@article{HenryLanglandsStraka2010,
  author  = {Henry, B. I. and Langlands, T. A. M. and Straka, P.},
  title   = {Fractional {F}okker-{P}lanck equations for subdiffusion with space- and time-dependent forces},
  journal = {Phys. Rev. Lett.},
  volume  = {105},
  pages   = {170602},
  year    = {2010},
  doi     = {10.1103/PhysRevLett.105.170602}
}

\end{document}